 \documentclass[final,5p,times,twocolumn,authoryear]{elsarticle}

\usepackage{amssymb}
\usepackage{url}
\usepackage{aas_macros}
\usepackage{color}
\newcommand{\intl}{\textit{INTEGRAL}}
\newcommand{\ib}{\textit{IBIS}}

\newcommand{\isg}{\textit{ISGRI}}

\def\cnts{cts s$^{-1}$ pix$^{-1}$}
\def\flux{erg s$^{-1}$ cm$^{-2}$}

\def\scw{\textit{ScW}}

\journal{High Energy Astrophysics}

\begin{document}

\begin{frontmatter}

%% Title, authors and addresses

%% use the tnoteref command within \title for footnotes;
%% use the tnotetext command for theassociated footnote;
%% use the fnref command within \author or \affiliation for footnotes;
%% use the fntext command for theassociated footnote;
%% use the corref command within \author for corresponding author footnotes;
%% use the cortext command for theassociated footnote;
%% use the ead command for the email address,
%% and the form \ead[url] for the home page:
%% \title{Title\tnoteref{label1}}
%% \tnotetext[label1]{}
%% \author{Name\corref{cor1}\fnref{label2}}
%% \ead{email address}
%% \ead[url]{home page}
%% \fntext[label2]{}
%% \cortext[cor1]{}
%% \affiliation{organization={},
%%            addressline={}, 
%%            city={},
%%            postcode={}, 
%%            state={},
%%            country={}}
%% \fntext[label3]{}

\title{The properties of the Galactic hard X-ray and soft $\gamma$-ray background based on 20 years of INTEGRAL/IBIS observations}

%% use optional labels to link authors explicitly to addresses:
%% \author[label1,label2]{}
%% \affiliation[label1]{organization={},
%%             addressline={},
%%             city={},
%%             postcode={},
%%             state={},
%%             country={}}
%%
%% \affiliation[label2]{organization={},
%%             addressline={},
%%             city={},
%%             postcode={},
%%             state={},
%%             country={}}

\author[first]{Roman Krivonos}
\author[first]{Ekaterina Shtykovskaya}
\author[first]{Sergey Sazonov}
\affiliation[first]{organization={Space Research Institute of the Russian Academy of Sciences},%Department and Organization
            addressline={Profsoyuznaya 84/32}, 
            city={Moscow},
            postcode={117997}, 
            state={},
            country={Russia}}

\begin{abstract}
%% Text of abstract
We present results of a study of the Galactic hard X-ray and soft $\gamma$-ray background emission performed with the \ib\ telescope aboard the \intl\ observatory using data obtained over more than 20 years of operations.  The study of the Galactic background at energies between 10 keV and a few MeV is problematic due to the contribution of point sources, high instrumental background and large-scale extent of the emission, which leads to the need of utilizing complex model-dependent methods. Using the unique properties of the \ib\ coded-mask telescope, we developed a model-independent approach to study diffuse continuum emission near the Galactic plane in the 25$-$60, 60$-$80, and 80$-$200~keV bands. The comparison of the 25$-$60~keV longitude profile with the near infrared intensity shows excellent agreement, confirming the stellar origin of the Galactic Ridge X-ray Emission (GRXE). The Galactic X-ray background is significantly detected from the direction of the Galactic bulge up to 200 keV. We built broad-band spectra of the Galactic background in three large regions, the Galactic bulge and two spiral arms at $l\approx\pm 20^{\circ}$. The spectral analysis reveals two distinct components with a minimum at about 80~keV. The low-energy ($\lesssim 60$~keV) component, associated with the GRXE, is consistent with a one-dimensional accretion flow model of intermediate polars with an average white dwarf mass of about $0.7$~$M_{\odot}$. The high-energy part of the spectrum, dominating above ${\sim}60$~keV and attributed to the $\gamma$-ray background, is consistent with a power-law model with photon index $\Gamma=1.55$. The total 30$-$80~keV flux budget of $1.5{\times}10^{-9}$\flux\ observed within the effective {\ib} field of view (${\approx}286$~deg$^{2}$) in the Galactic bulge region, consists of $2/3$ of GRXE and $1/3$ of $\gamma$-ray background. Finally, we provide the Python code of the \ib/\isg\ background model, which can be used to measure the X-ray intensity of the Galactic background in different parts of the Milky Way.
\end{abstract}

%%Graphical abstract
%\begin{graphicalabstract}
%\includegraphics{grabs}
%\end{graphicalabstract}

%%Research highlights
%\begin{highlights}
%\item Research highlight 1
%\item Research highlight 2
%\end{highlights}

\begin{keyword}
%% keywords here, in the form: keyword \sep keyword, up to a maximum of 6 keywords
X-rays: general \sep Galaxy: bulge \sep Galaxy: general \sep gamma rays: diffuse background \sep X-rays: diffuse background

%% PACS codes here, in the form: \PACS code \sep code

%% MSC codes here, in the form: \MSC code \sep code
%% or \MSC[2008] code \sep code (2000 is the default)

\end{keyword}

\end{frontmatter}

%\tableofcontents

%% \linenumbers

%% main text

\section{Introduction}

%Observed in different energy bands the background emission of the Galaxy are related with different processes. For example, high energy 10--100~MeV background emission are caused by interactions of high energy cosmic rays with interstellar matter in Galaxy. This emission is a composition of a large amount of active galactic nuclei (AGN) \citep{}. 

The stellar origin of the Galactic ridge X-ray emission (GRXE) has been strongly supported by a morphological study with the RXTE observatory in the pioneering work by \cite{2006A&A...452..169R}, who demonstrated that the 3$-$20~keV map of the GRXE closely follows the near-infrared brightness distribution of the Galaxy and thus traces the Galactic stellar mass distribution. \cite{2006A&A...452..169R} also predicted a high-energy cut-off in the GRXE spectrum above 20 keV, which was later discovered with \intl\ \citep{2007A&A...463..957K, 2010A&A...512A..49T,2008ApJ...679.1315B,2011ApJ...739...29B,2022A&A...660A.130S}, Suzaku \citep{2012ApJ...753..129Y} and NuSTAR \citep{2019ApJ...884..153P}. The GRXE is associated with the old stellar population of the Galaxy, mostly with the hard X-ray emission from accreting white dwarfs, in particular polars and intermediate polars, and with the softer emission from coronally active stars. The accretion column on the magnetic poles of accreting white dwarfs emits optically thin thermal X-ray emission. The integrated emission of a large number of such faint Galactic X-ray sources constitutes the bulk of the observed GRXE \citep[see e.g.,][]{2006A&A...450..117S,Revnivtsev_2008,2009Natur.458.1142R,Revnivtsev_2011,2020NewAR..9101547L}.

In contrast to GRXE, the Galactic diffuse continuum emission (GCDE) at energies above 100~keV is believed to be truly diffuse in origin. GCDE at energies below $\approx$70~MeV is mainly due to the interactions of cosmic-rays (CR) with the interstellar material and radiation fields \citep{1972ApJ...177..341K,1978ApJ...225..591K,1997ApJ...481..205H}. The Compton scattering of the CR leptons with the interstellar radiation field and the cosmic microwave background is the major contribution to the GCDE \citep{2005ICRC....4...77P,2008ApJ...682..400P}. 
Observations of Galactic $\gamma$-ray diffuse emission thus provide a unique opportunity to study the properties of Galactic CR particles \citep{2011ApJ...739...29B,2022A&A...660A.130S,2023ApJ...959...90K}. 

However, the detection of high-energy Galactic emission has been a notoriously difficult problem. The signal is very weak and thus strongly depends on the accuracy of the instrumental background subtraction. A number of detections and non-detections of high-energy Galactic emission have been claimed \cite[e.g.][]{1985ApJ...294L..13R,2001ApJ...559..282K,2005A&A...444..495S,2008ApJ...679.1315B}. Another difficulty is the unknown spatial distribution of cosmic-ray induced background, which leads to the need of using complex model-dependent methods \citep[e.g.,][]{2011ApJ...739...29B,2022A&A...660A.130S}. Therefore, any independent study of the Galactic background in hard X-ray and soft $\gamma$-ray bands is important to improve our understanding of the energetic content of the Milky Way.

Thanks to the unique combination of properties of INTEGRAL's \ib\ and SPI telescopes -- broad-band coverage, large fields of view (FOV) and possibilities to eliminate the contribution of discrete sources from the total measured photon flux, the study of the Galactic hard X-ray and soft $\gamma$-ray background is possible. The aim of this work is to provide model-independent measurements of the wide-angle morphology and broad-band spectrum of the Galactic background emission above 25~keV up to 200~keV with the {\ib} telescope. 

The paper is organized as follows. Section~\ref{sec:data} describes the multi-year \intl\ all-sky observations and initial data reduction. Section~\ref{sec:coded} presents the concept of using the \ib\ telescope as a collimated instrument for studying the Galactic large-scale X-ray emission. The details of the method, namely the description of the \ib\ detector background model; flux calibration with the Crab Nebula; and the validation of the entire method using Sco~X-1 data, can be found in \ref{sec:bkg}; \ref{sec:calibr}; and \ref{sec:scox1}, respectively. The results are presented in Section~\ref{sec:res}, where flux measurement in different Galactic regions (Section~\ref{sec:flux}); longitude morphology in different energy bands up to 200 keV (Section~\ref{sec:morphology}); and spectral analysis (Section~\ref{sec:spec}) of Galactic emission can be found.

%\cite{2010A&A...512A..49T}: The derived spectrum of the GRXE confirms the presence of a minimum around 80~keV with improved statistics and yields an estimate of ${\sim}0.6$~Msun for the average mass of white dwarfs in the Galaxy.

\section{Observations and data reduction}
\label{sec:data}

For this work we use all publicly available \intl\ \citep{2003A&A...411L...1W} data acquired with the \ib\ coded-aperture telescope \citep{2003A&A...411L.131U} from May 2003 to February 2024 (orbits 70$-$2740). We considered only the data of the \isg\ detector \citep{2003A&A...411L.141L}, which provides data in the energy band 17$-$1000~keV with high sensitivity in the transition interval from hard X-rays to soft $\gamma$-rays (17$-$200~keV).

We reduced \ib/\isg\ data with a proprietary analysis package developed at IKI\footnote{Space Research Institute (IKI), Moscow,
Russia}  \citep[see details in][]{2010A&A...519A.107K,2012A&A...545A..27K,2014Natur.512..406C}. Below we describe details of the data analysis that are relevant for the current work.

We first applied the energy calibration for the registered IBIS/ISGRI detector events with the ``Off-line Scientific Analysis'' (\texttt{OSA}) software package provided by the \intl\ Science Data Centre (ISDC) for Astrophysics\footnote{\url{https://www.isdc.unige.ch/integral}} up to the COR level. We used  the latest \texttt{OSA} version 11.2, which provides energy calibration for the whole period of observations (however, see \ref{sec:calibr} for additional details).  

We screened the {\intl} data before subsequent analysis to reduce systematic noise and remove all contaminated observations and observations with insufficient statistics. Hereafter, by an {\intl} observation we mean an individual ``science window'' ({\scw}), a continuous time interval during which all data acquired by the {\intl} instruments result from a specific attitude orientation state. A typical duration of a {\scw} is about 2~ks. If some {\scw} did not satisfy all the imposed criteria, it was skipped. We screened all {\scw}s near the beginning and end of \intl\ revolution with orbital phases ${<}0.2$ or ${>}0.8$, due to increased background near the radiation belts; the data when \ib\ was operated not in its main regime (modes 41 and 43); and {\scw}s with exposure times less then $700$~s. As a result, 131440 {\scw}s satisfies the selection criteria, with the total (dead-time corrected) exposure of 226~Ms. 

\subsection{Sky regions}

We defined three wide Galactic regions to measure the X-ray flux and to extract broad-band spectra of the Galactic Background, as listed in Table~\ref{tab:skyreg}. The first region effectively includes X-ray emission from the Galactic Bulge (GB), the inner few kpc of our Galaxy \citep{1995ApJ...445..716D,Zoccali_2016}. The two nearby, ``L$+$20'' and ``L$-$20'', regions approximately cover the Scutum and Norma Galactic spiral arms, respectively. Both of these regions approximately cover the structure of the Galactic disk at $|l|<50^{\circ}$, which contains the main stellar mass component of the Galaxy. We additionally defined a high-latitude 3C 273 and the Coma cluster region to validate the analysis procedure. This region has been the target of the deepest exposure with {\intl} extragalactic surveys \citep{2005ApJ...625...89K,2008A&A...485..707P,2016MNRAS.459..140M}.

To better describe the large-scale morphology of the Galactic Background, we additionally defined narrow Galactic longitude intervals. As we here use the \ib/\isg\ telescope, with a FOV of ${\sim}15^{\circ}\times15^{\circ}$ (FWHM), as a collimated instrument, the angular resolution of our method is approximately ${\sim}15^{\circ}$. Based on that, we defined longitude bins with a width of 15$^{\circ}$ and latitude height of $|b|<10^{\circ}$.

\subsection{Energy bands}

Due the loss of \isg\ sensitivity at low energies caused by the ongoing detector degradation, we define our energy bands above $25$~keV. To characterize the Galactic X-ray background, we use three wide energy bands. First, the lower band 25$-$60~keV is chosen to better fit the spectrum of the GRXE, which has a cutoff at energies ${\sim}30-50$~keV due to the typical cutoff in spectra of magnetic CVs \citep{2005A&A...435..191S}. Second, the intermediate 60$-$80~keV band is selected as the interval where the transition between the GRXE and the $\gamma$-ray background is observed. Third, the 80$-$200~keV band is selected as dominated by the contribution of the $\gamma$-ray background. 

For the spectral analysis, we selected 21 logarithmically spaced energy bands between 25 and 185~keV. This spectral binning provides the necessary statistics and a suitable energy resolution to describe the two main continuum components in the broad-band spectrum. For spectral fitting, we use a diagonal energy redistribution matrix that reproduces the Crab-like spectrum, represented as $10.0\times E^{-2.1}$~keV~photons~cm$^{-2}$~s$^{-1}$~keV$^{-1}$ \citep[see, e.g.][]{2007A&A...467..529C}.

%For our analysis, we used individual sky images ('science window' or ScW) obtained for each \intl\ observation in the following energy bands: 17--22, 22--29, 29--37, 37--49, 49--63, 63--82, 82--107, 107--139, 139--180, 180--234, 234-305, and 17--60 keV. 

\begin{table*}
\centering
\caption{Description of sky regions, statistics of available observations and measured X-ray flux of Galactic X-ray background.}
\label{tab:skyreg}
\begin{tabular}{c|r|r|c|l|c|c|c|c}
\hline
Name & Lon & Lat & Size$^{\rm a)}$ & {\scw}s & Exp. & \multicolumn{3} {c} {Flux$^{\rm b)}$ (mCrab FOV$^{-1}$)} \\
 & deg. & deg. &  &  & Ms. & 25$-$60 keV & 60$-$80 keV & 80$-$200 keV \\
\hline
GB & 0$^{\circ}$ & 0$^{\circ}$ & $6^{\circ}\times6^{\circ}$ & 6697 & 9.8 & $154.9\pm3.1$ & $96.5\pm9.2$ & $102.2\pm13.3$ \\
L$+$20 & $+$20$^{\circ}$ & 0$^{\circ}$ & $20^{\circ}\times10^{\circ}$ & 3160 & 4.3 & $69.1\pm2.3$ & $23.0\pm5.2$ & $56.7\pm9.8$ \\
L$-$20 & $-$20$^{\circ}$ & 0$^{\circ}$ & $20^{\circ}\times10^{\circ}$ & 5068 & 7.7 & $67.9\pm1.7$ & $33.3\pm5.3$ & $29.3\pm9.8$\\
3C 273/Coma & $-$70$^{\circ}$ & 70$^{\circ}$ & $40^{\circ}\times40^{\circ}$ & 5400 & 11.7 & $-1.0\pm0.7$ & $1.6\pm2.0$ & $5.6\pm3.4$ \\
%LON$+$40 & $+$43$^{\circ}$ & 0$^{\circ}$ & $20^{\circ}\times10^{\circ}$ & 4577 & 7.6~Ms \\
\hline
\end{tabular}
\begin{flushleft}
$^{\rm a)}$ The centers of {\intl} {\scw}s are selected within the rectangular regions with specified size. Note that the {\ib} FOV, used as a collimated instrument is ${\sim}15^{\circ}\times15^{\circ}$ (FWHM).  \\
$^{\rm b)}$ The flux is measured within {\ib} FOV with effective solid angle $\Omega\approx286$~deg$^{2}$.
\end{flushleft}
\end{table*}

\section{\ib\ telescope coding aperture}
\label{sec:coded}

The coded-aperture paradigm implemented for the \ib\ telescope does not allow one to directly investigate extended structures that are significantly greater than the pixel size. However, \cite{2007A&A...463..957K} showed that the \ib\ telescope can be used for GRXE studies as a collimated instrument, which collects emission from both point sources and the Galactic background, and the latter can be measured separately after taking the internal detector background into account.

To work in collimated mode, we do not need to visualize the sky for each {\scw}, i.e. to apply the standard coded-aperture image reconstruction. However, we iteratively subtract shadowgrams illuminated by point sources from the detector plane using the known mask pattern \citep[see details in ][]{2010A&A...519A.107K}.  This procedure is performed for each {\scw}. The list of X-ray sources is taken from the recent 17-year \intl/\ib\ all-sky survey  \citep{2022MNRAS.510.4796K}. Hereafter we consider only `clean' \isg\ detector count rate after removal of the contribution of point-like X-ray sources. 

Note that the function of the {\ib} mask pattern is not ideally known \citep{Soldi:20137E}, so that the source removal procedure is not perfect. A small fraction of source flux can be left or oversubtracted on the detector \citep{2007A&A...475..775K}. This effect becomes significant for very bright sources. For this reason, we additionally screened the data in each energy band for strong X-ray transients manually or using an iterative $4\sigma$ clipping algorithm. After this step, the {\isg}  detector count rate in each {\scw} contains three components: 
\begin{enumerate}
    \item Cosmic X-ray background \citep[CXB, see e.g.][]{2003A&A...411..329R,2007A&A...467..529C,2021MNRAS.502.3966K};
    \item Galactic X-ray background (GXB), if the telescope's FOV is directed towards the Galactic plane; Since we work in the energy range from 25 to 200~keV, the GXB contains both the hard X-ray background (i.e. GRXE) below ${\sim}60$~keV and the Galactic soft $\gamma$-ray background, which dominates above 60~keV;
    \item detector internal background, caused by different processes including activation of different elements of the spacecraft, interaction of the detector material with cosmic-rays, etc. \citep{2003A&A...411L.167T}.
\end{enumerate}

The measurement of the surface brightness of the GXB subtended by the \ib\ FOV is defined as the difference between the observed detector count rate for each {\scw} and the count rate predicted by the background model (\ref{sec:bkg}), and divided by the smooth polynomial function of the observed Crab count rate (\ref{sec:calibr}). The resulting value is, therefore, expressed in Crab (or mCrab) units, which can be converted to physical units (\flux) using the Crab spectral model (\ref{sec:calibr}).

%Also, the data, which was contaminated by different types of bursts (X-ray bursts, insufficient subtraction of bright sources and X-ray transients, interactions with heavy cosmic ray particles, etc.), was selected using the {\sc sigma\_clip} tool a part of {\sc Astropy v4.3.1} package in Python (version 3.8), which filters the data by a standard deviation ($<=4\sigma$) from a center value. After the data clipping, the residual intervals with the contamination were selected manually. Each energy band was filtered individually. Finally, obvious outliers were manually trimmed from the remaining data. 

%The possibility of the removing of the point source contribution from the full detector count rate allow us to obtain the GRXE and background count rates. The \ib\ background count rate contains the X-ray background count rate due to interactions of the Interstellar medium with cosmic rays and instrumental background, which is caused by different processes like as: activation of different parts of the spacecraft, interaction of the detector matter with cosmic rays etc. Therefore, an accuracy of the GRXE measurements directly depends on a background estimation accuracy. 

% \textbf{Then we extracted count rates of source-free images (ScWs without detected point sources), which were used for the background model determination and the GRXE signal .}

%\input{tables/ebands}

%\subsection{Data filtering}
%\label{sec:filt}

%The regions close to the known bright X-ray sources were removed from the analysis. Radii of the removed regions and source names are in Table~\ref{tab:src}.

\section{Results}
\label{sec:res}
To give a large-scale picture, below we first present the measurement of the Galaclic X-ray Background intensity in wide sky regions and broad energy intervals. Then we demonstrate its longitudinal morphology and provide a broad-band spectral analysis.

To measure the X-ray surface brightness per {\ib} FOV in a given sky region and a given energy band, we run a Gaussian fitting procedure to estimate the best-fitting mean flux value and its error, as described in \ref{sec:bkg}.

\subsection{Flux measurement}
\label{sec:flux}

Table~\ref{tab:skyreg} presents the flux values obtained by this approach for the GB and L$\pm$20 spiral arms regions. The Galactic X-ray background is significantly detected in the GB over all energy bands up to 200 keV. In the 25$-$60~keV band, the GRXE flux in the spiral arms is a factor of two lower than in the GB, which is consistent with the NIR intensity distribution convolved with the \ib/\isg\  collimator response function \citep[see Fig.~11 in][]{2007A&A...463..957K}. In the intermediate 60$-$80 keV and $\gamma$-ray band 80$-$200~keV, the measured flux in the spiral arms is detected at $3-6\sigma$ confidence.

\subsection{Morphology}
\label{sec:morphology}

\begin{figure}
	\includegraphics[width=\columnwidth]{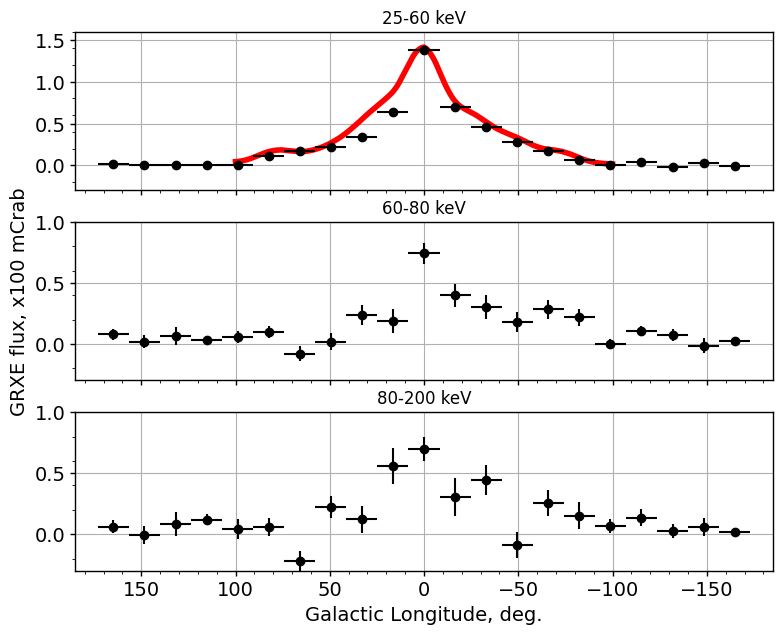}
    \caption{Galactic longitude profiles of Galactic X-ray Background in three energy bands. The red solid line represents the intensity profile of the Galactic NIR emission measured by COBE/DIRBE at 4.9 µm. The NIR map was convolved with the IBIS collimator response.}
    \label{fig:ridge:galprof}
\end{figure}

We constructed a longitude profile of the Galactic background emission in broad energy bands. \intl\ {\scw}s within each Galactic longitude interval were used to estimate the X-ray flux as described in Sect.~\ref{sec:flux}. Figure~\ref{fig:ridge:galprof} shows the longitude profiles for the three wide energy bands.

We constructed the map of the Galaxy in the near infrared (NIR) spectral band at 4.9~${\mu}$m using data of COBE/DIRBE observations, as described in \cite{2007A&A...463..957K}. The map of the NIR intensity was convolved with the \ib/\isg\ collimator response function. The Galactic longitude profile of the COBE/DIRBE NIR intensity is shown by the solid line in Fig.~\ref{fig:ridge:galprof} for the 25$-$60~keV band. Note that NIR intensity was arbitrary scaled to approximately match the X-ray data. As seen from the figure, the 25$-$60~keV intensity distribution closely follows the NIR intensity and thus traces the stellar mass density in the Galaxy.

%The longitude profile of the Galactic ridge emission (Fig.~\ref{fig:ridge:galprof}) shows possible imprint of the disk spiral structure. As noticed by \cite{2006A&A...452..169R} who analysed RXTE data in 3$-$20~keV band, the ridge emission appears stronger at $l=-60^{\circ}$ than at $l=60^{\circ}$, which is probably caused by a significant enhancement of the GRXE in spiral arms. 

\subsection{Spectrum of the Galactic X-ray background}
\label{sec:spec}

\begin{figure}
	\includegraphics[width=\columnwidth]{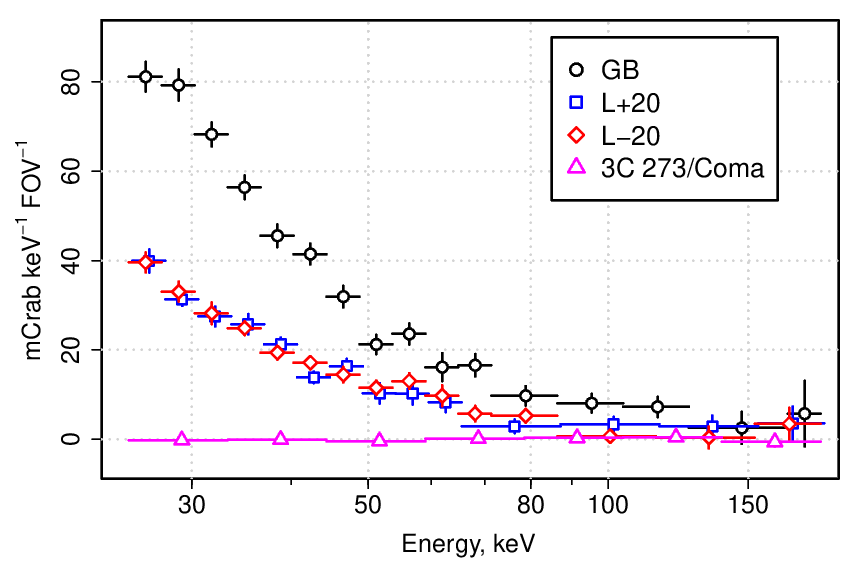}
    \caption{The X-ray spectra in mCrab units per \ib\ FOV observed in different sky regions (Table~\ref{tab:skyreg}). The L$+$20 and L$-$20 spectra are shifted with respect to each other for better visibility.}
    \label{fig:ridge:mcrab}
\end{figure}

In Fig.~\ref{fig:ridge:mcrab}, we show the raw X-ray spectra of the Galactic X-ray Background in mCrab units (see Sect.~\ref{sec:flux}) in the sky regions defined in Table~\ref{tab:skyreg}. The brightest intensity is observed, as expected, in the GB region. The X-ray intensity of the spiral arms (L$+$20 and L$-$20) are similar to each other and a factor of ${\sim}2$ less than in the GB. For comparison, the measured X-ray intensity of the extragalactic 3C~273/Coma field is consistent with zero.

The spectral shape of the Galactic X-ray Background is the main source of information about its composition. A stellar origin is expected at energies below $\sim 60$ keV (GRXE), while the $\gamma$-ray background is expected to dominate above 80~keV. 

\begin{figure}
	\includegraphics[width=\columnwidth]{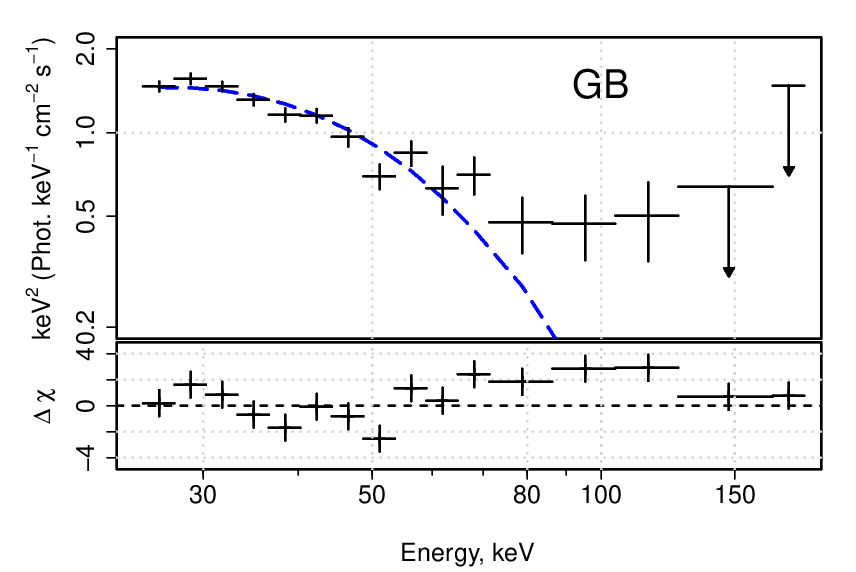}
    \caption{Spectrum of the GXB in the GB region, approximated by a power law with a high-energy cut-off (in dashed blue). For better plotting, no more than three adjacent spectral bins are combined to have $3\sigma$ detection. The upper limits are shown at the level of $2\sigma$. The spectral flux is measured within the \ib\ FOV with effective solid angle of $\Omega\approx286$~deg$^{2}$.}
    \label{fig:solo}
\end{figure}

First, we approximated the 25$-$185~keV spectrum of the GB region by a single power law with a fixed slope $\Gamma^{\rm cut}=0$, free high-energy cut-off energy $E_{\rm cut}$ and free normalization (model \texttt{cutoffpl} in XSPEC notation), to represent the GRXE, as shown in Fig.~\ref{fig:solo}. This single-component model\footnote{A similar fit with the polar model described below gives $\chi^{2}_{\rm red}=1.44$ for 19 d.o.f. and M$_{\rm WD}=0.85\pm0.06$~M$\odot$.}, with $E_{\rm cut}{\approx}13$~keV, poorly describes the data, with $\chi^{2}_{\rm red}=2.36$ for 19 d.o.f., mainly due to the evident presence of a high-energy excess above ${\sim}60$~keV. To account for this excess, we added a second power-law component with slope $\Gamma^{\rm pow}=1.55$, attributed to the $\gamma$-ray background, which significantly improved the fit: $\chi^{2}_{\rm red}=0.89$ for 18 d.o.f. The values $\Gamma^{\rm cut}=0$ and $\Gamma^{\rm pow}=1.55$ for this analysis were adopted from previous studies, namely according to the \intl/SPI study of the Galactic diffuse emission by \cite{2008ApJ...679.1315B}, also consistent with the results of later works \citep{2010A&A...512A..49T,2011ApJ...739...29B,2022A&A...660A.130S}.

\begin{figure}
\includegraphics[width=\columnwidth]{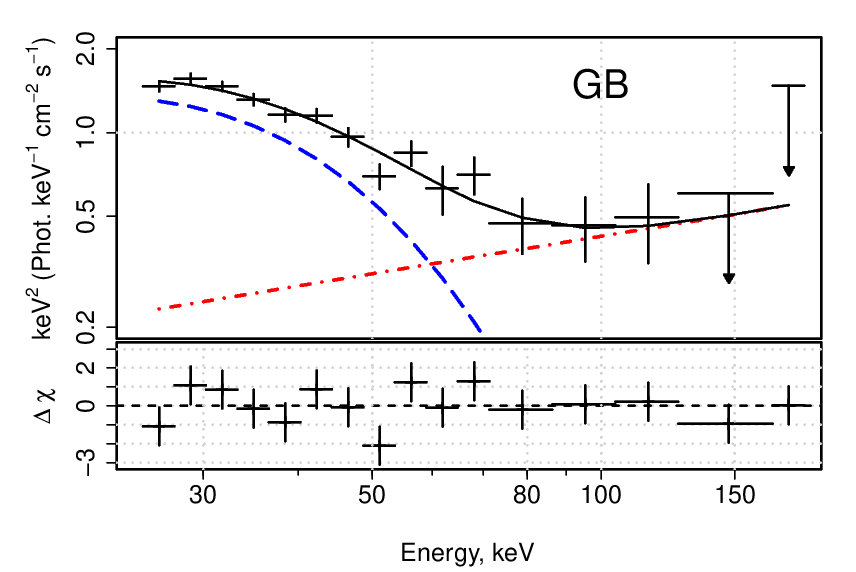}
\includegraphics[width=\columnwidth]{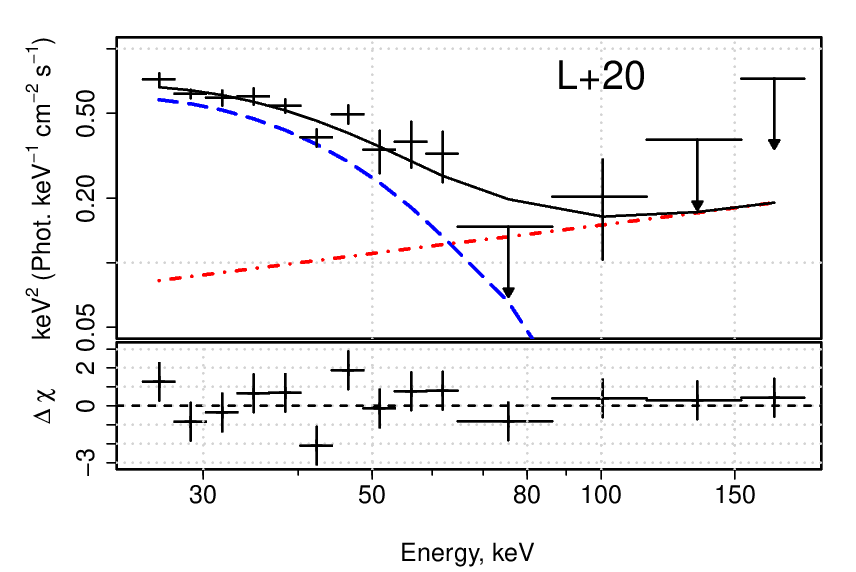}
\includegraphics[width=\columnwidth]{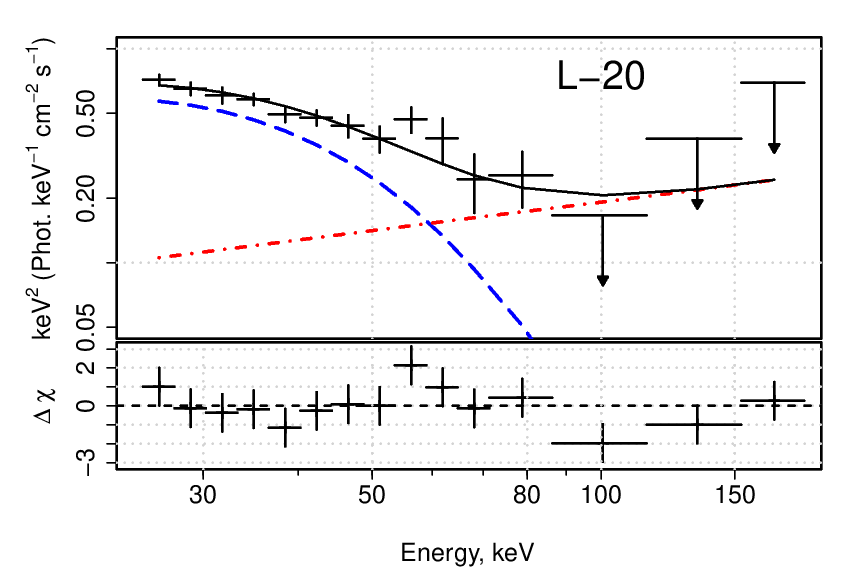}
    \caption{Spectra of the GXB in different Galactic regions (Table~\ref{tab:skyreg}). The spectral modeling is done with a combination of a power law with a high-energy cut-off (\texttt{cutoffpl}, in dashed blue) and a power law with $\Gamma$=1.55 (dash-dotted red). For detailed description also see Fig.~\ref{fig:solo}.}
    \label{fig:ridge:spec}
\end{figure}

We next ran a spectral fitting procedure for all three Galactic regions with the following free parameters: the normalization of both power-law components and the high-energy cut-off energy $E_{\rm cut}$. The normalization of each component was estimated with the \texttt{cflux} model in \texttt{xspec} in the 30$-$80~keV energy band. The best-fitting parameters are listed in Table~\ref{tab:spec}. This spectral model gives a good description of the data with $\chi^{2}_{\rm red} = 0.9-1.5$ for 18 d.o.f. Figure~\ref{fig:ridge:spec} shows the X-ray spectra and best-fit spectral models for the three Galactic regions.

\begin{table}
\centering
\caption{Best-fitting parameters of the model \texttt{cutoffpl+pow}. }
\label{tab:spec}
\begin{tabular}{l|l|l|l}
\hline
Parameter & GB & L+20 & L-20 \\ % & LON+40\\
%\hline
%Sys. error  & 5\%  & 0\% & 0\% & 15\% \\
\hline
$\Gamma^{\rm cut}$ (fixed) & 0.0 & 0.0 & 0.0 \\ % & 0.0\\
$E_{\rm cut}$, keV & $11.2\pm0.9$ & $11.2\pm 1.5$ &  $11.2\pm1.7$ \\ % & $11.0_{-2.8}^{+3.2}$ \\
$F_{\rm 30-80\ keV}^{\rm cut}$$^{\rm a)}$  & $9.7\pm1.4$ & $4.3_{-1.0}^{+1.2}$ & $4.3_{-0.9}^{+1.1}$ \\ % & $1.5\pm0.2$ \\

$\Gamma^{\rm pow}$ (fixed) &  1.55 & 1.55 & 1.55 \\ % & 1.55 \\
$F_{\rm 30-80\ keV}^{\rm pow}$$^{\rm a)}$  & $4.9\pm 1.4$ & $1.7\pm 1.3$ & $2.2_{-1.1}^{+1.0}$ \\ % & $<0.4$ \\

$F_{\rm 30-80\ keV}^{\rm total}$$^{\rm a)}$  & $14.6\pm 0.2$ & $6.1\pm0.1$ & $6.5\pm 0.1$ \\ % & $<0.4$ \\

$F_{\rm 30-80\ keV}^{\rm cut}$/$F_{\rm 30-80\ keV}^{\rm total}$ & $0.66\pm 0.10$ & $0.70\pm 0.57$ & $0.66\pm 0.37$ \\ 
$F_{\rm 30-80\ keV}^{\rm pow}$/$F_{\rm 30-80\ keV}^{\rm total}$ & $0.33\pm 0.10$ & $0.27\pm 0.20$ & $0.34\pm 0.14$\\

$\chi^{2}_{\rm red}$/dof &  0.89/18 & 1.47/18 & 1.13/18 \\ % & 1.43/18 \\
\hline
\end{tabular}
\begin{flushleft}
$^{\rm a)}$ The flux is expressed in units $10^{-10}$\flux~FOV$^{-1}$    
\end{flushleft}
\end{table}

To describe the hard X-ray emission below $\sim 60$~keV with a physically motivated model, we use a one-dimensional accretion flow model that accounts for the density and temperature profile of the accretion column of intermediate polars (IPs) developed by \cite{2005A&A...435..191S}, which are the dominant contributors in hard X-rays. We use this IP model (IPM) to derive the average WD mass implied by our GB, L$+$20 and L$-$20 spectra. The IPM model has two parameters: $M_{\rm WD}$ and normalization. By repeating the fitting procedure as described above, we estimated optimal model parameters, which are listed in Table~\ref{tab:spec:polar}. The estimated average WD masses for all Galactic regions are consistent with each other within the uncertainties and in good agreement with previous estimates: $M_{\rm WD}{\approx}0.5-0.66$~M$_{\odot}$ \citep{2007A&A...463..957K, 2010A&A...512A..49T, 2012ApJ...753..129Y,2013MNRAS.428.3462H,2019ApJ...884..153P}.

\begin{table}
\centering
\caption{Best-fitting parameters of the model \texttt{IPM+pow}. }
\label{tab:spec:polar}
\begin{tabular}{l|l|l|l}
\hline
Parameter & GB & L+20 & L-20 \\ % & LON+40\\
%\hline
%Sys. error  & 5\%  & 0\% & 0\% & 15\% \\
\hline
$M_{\rm WD}$, M$_{\odot}$ & $0.70\pm 0.09$ & $0.73\pm0.14 $ &  $0.80\pm 0.15$ \\ % & $9.8\pm0.9$ \\
$F_{\rm 30-80\ keV}^{\rm IPM}$$^{\rm a)}$  & $10.7_{-1.7}^{+1.9}$ & $5.1\pm 1.4$ & $5.6\pm 1.5$ \\ % & $1.5\pm0.2$ \\

$\Gamma^{\rm pow}$ (fixed) &  1.55 & 1.55 & 1.55 \\ % & 1.55 \\
$F_{\rm 30-80\ keV}^{\rm pow}$$^{\rm a)}$  & $4.0\pm 1.7$ & $1.1_{-1.1}^{+1.4}$ & $1.1_{-1.1}^{+1.4}$ \\ % & $<0.4$ \\

$F_{\rm 30-80\ keV}^{\rm total}$$^{\rm a)}$  & $14.7\pm 0.2$ & $6.2\pm 0.1$ & $6.7\pm 0.1$ \\ % & $<0.4$ \\

$F_{\rm 30-80\ keV}^{\rm IPM}$/$F_{\rm 30-80\ keV}^{\rm total}$ & $0.73\pm 0.13$ & $0.82\pm 0.23$ & $0.83\pm 0.22$ \\ 
$F_{\rm 30-80\ keV}^{\rm pow}$/$F_{\rm 30-80\ keV}^{\rm total}$ & $0.27\pm 0.11$ & $0.17\pm 0.22$ & $0.16\pm 0.20$\\

$\chi^{2}_{\rm r}$/dof &  0.76/18 & 1.27/18 & 0.83/18 \\ % & 0.09/18 \\
\hline
\end{tabular}
\begin{flushleft}
$^{\rm a)}$ The flux is expressed in units $10^{-10}$\flux~FOV$^{-1}$    
\end{flushleft}
\end{table}

%The background model (Sect.~\ref{sec:bkg}) predicts the detector count rate for any observation in \texttt{GAL} region containing Galactic X-ray Background. The residual count rate is converted to mCrab units using {\bf the central value} of the evolving \ib/\isg\ collimator response function (Sect.~\ref{sec:calibr}). As a result, we have measurement of Galactic X-ray Background for each \textit{ScW} within \texttt{GAL} region for each energy band.

%Чтобы получить карты неба, мы должны сначала получить чистый калиброванный сигнал. Для этого из данным, извлеченным из региона GRXE, была применена энергетическая калибровка, а затем вычтена модель фона, полученная в предыдущем параграфе. Полученные данные были отфильтрованы в соответствии с процедурой, описанной в секции~\ref{sec:filt}. Таким образом были получены чистые калиброванные значения потоков, накопленных за каждое наблюдение GRXE. 

In contrast to our previous analysis, based on a much smaller \isg/\ib\ data set \citep{2007A&A...463..957K}, when the measured spectrum of the GXB was characterized by a high-energy cutoff at ${\sim}50$~keV, we have now detected significant emission in the intermediate 60$-$80~keV band and the derived spectrum reveals a minimum at about 80 keV, as also found in other studies \citep{2010A&A...512A..49T,2008ApJ...679.1315B,2012ApJ...753..129Y}. We attribute the high-energy cut-off reported by \cite{2007A&A...463..957K} to low statistics and unaccounted systematics, which resulted in an underestimation of the GRXE temperature and, consequently, of the average mass of accreting WDs ($\sim 0.5$~$M_{\odot}$). The current analysis implies a higher WD mass $\sim 0.7$~$M_{\odot}$, in better agreement with other works.

To validate the ability of our method to correctly extract broadband spectral information, ideally it would be good to have an extended object with a known spectrum. Unfortunately, there are no such objects on the sky, except for the GXB itself (the uniform CXB is not suitable for this purpose). Qualitatively, the correctness of the method is indicated by the inferred similarity of the GXB spectral shapes in the L$+$20 and L$-$20 regions, which are spatially separated from each other. For a more detailed test, we can also use a spectrum of a known bright point source obtained in the {\ib}/{\isg} collimator mode, which is done in \ref{sec:scox1}, using Sco~X-1 data. We demonstrated that the time-averaged spectrum of Sco~X-1 is consistent with the {\ib}/{\isg} spectrum observed by \cite{2014MNRAS.445.1205R} in point-source mode, i.e. by using the standard {\ib} sky reconstruction method.

\section{Summary and conclusions}

We studied the large-scale morphology and spectral properties of the Galactic hard X-ray and soft $\gamma$-ray background emission from 25 to 200 keV using multi-year \intl\ data. The aim of this work is to provide a model-independent measurement of large-scale extended emission of the Milky Way in the problematic domain of observations. We developed a continuously-calibrated \ib/\isg\ detector background model, characterized by ${\sim}1-2\%$ systematic uncertainty, which allowed us to measure the distribution of X-ray intensity within the Galactic plane ($|l|<90^{\circ}$, $|b|<30^{\circ}$).

In the hard X-ray band 25$-$60~keV, the Galactic background is significantly detected in the region of the Galactic bulge at the level of ${\sim}155$~mCrab per \ib\ FOV, represented by effective solid angle of  $286$~deg$^{2}$. The measured flux in the Galactic disk is a factor of two less of this value, which is consistent with the NIR intensity distribution. A more detailed comparison of the 25$-$60~keV longitude profile with NIR intensity shows excellent agreement, indicating a stellar origin of the GRXE, consistent with previous studies.

In the intermediate 60$-$80 keV band, the Galactic background emission is significantly ($10\sigma$) detected in the bulge region with flux ${\sim}100$~mCrab~FOV$^{-1}$. The emission of the Galactic disk is measured at the level of ${\sim}30$~mCrab~FOV$^{-1}$ with 4$-$6$\sigma$ confidence. The longitude morphology within $|l|<50^{\circ}$ is not peaked as NIR intensity and appears flattened, indicating a different origin, most likely the growing contribution of the cosmic-ray induced $\gamma$-ray background.
%, traced by Galactic neutral hydrogen (HI), and molecular gas (CO emission).

In the soft $\gamma$-ray band 80$-$200~keV, the Galactic emission is detected in the bulge at the level of ${\sim}100$~mCrab~FOV$^{-1}$ ($8\sigma$). The emission of the Galactic disk at $|l|<50^{\circ}$ is detected with flux ${\sim}30-60$~mCrab~FOV$^{-1}$ at low confidence (3$-$6$\sigma$). 

%{\bf drawn from large systematic uncertainties shown in longitude profile.}

The spectral analysis at energies between 25 and 185~keV reveals two distinct spectral components with a minimum at about 80 keV, as previously observed by different experiments. The low-energy component below ${\sim}$50$-$60~keV, coming from the GRXE, is well described by a power-law model $\Gamma=0$ with a high-energy cut-off $E_{\rm cut}\approx 11$~keV, and consistent with a one-dimensional accretion flow model of intermediate polars with an average WD mass of $M_{\rm WD}{\approx}0.7$~M$_{\odot}$. The high-energy part of the spectrum, attributed to $\gamma$-ray background at $E{\gtrsim}80$~keV is consistent with a power-law model characterized by the photon index $\Gamma=1.55$, as measured with early \intl/SPI studies. The total 30$-$80~keV flux budget observed with \ib/\isg\ in the region of the Galactic center $(14.7\pm 0.2)\times10^{-10}$\flux~FOV$^{-1}$, consists of ${\sim}2/3$ of GRXE and ${\sim}1/3$ of $\gamma$-ray background for the \texttt{cutoffpl} model (Table~\ref{tab:spec}). The same proportion is observed in the Galactic disk within the uncertainties. For the \texttt{IPM} model, the GRXE fraction in the total flux budget is higher (${\sim}80\%$, Table~\ref{tab:spec:polar}).

Finally, we provide the code of the \ib/\isg\ background model used in this work as a Python module available in a Git repository hosted by IKI\footnote{The package can be cloned from the repository by the command \texttt{git clone http://heagit.cosmos.ru/integral/ridge.git}. The installation and usage instructions are located inside the package and at the  \url{http://heagit.cosmos.ru} website.}. The package contains \isg\ detector count rate cleaned from the contribution of X-ray point sources, for each {\scw} in the range of the \intl\ orbits 70$-$2740 and in different energy bands, used in this work. This data set and code can be used to calibrate the \isg\ background model and measure the X-ray intensity of the Galactic background in different parts of the Milky Way.

\section*{Acknowledgments}
Dedicated to Mikhail Revnivtsev (1974-2016). We thank the anonymous referee for valuable comments. This work is based on observations with \intl, an ESA project with instruments and the science data centre funded by ESA member states (especially the PI countries: Denmark, France, Germany, Italy, Switzerland, Spain), and Poland, and with the participation of Russia and the USA. The authors are grateful to E.~M.~Churazov, who developed the INTEGRAL/IBIS data analysis methods and provided the software. This work was financially supported by grant 24-22-00212 from the Russian Science Foundation.

%% The Appendices part is started with the command \appendix;
%% appendix sections are then done as normal sections
\appendix

\section{\isg\ detector background model}
\label{sec:bkg}

The aim of this work is to estimate the flux of the Galactic hard X-ray and soft $\gamma$-ray astrophysical background, which is in fact the difference between the observed total \isg\ source-free detector count rate and the predicted internal background when \ib\ telescope is pointed towards the Milky Way. The contribution of the CXB is considered as a constant addition to the internal detector background \citep[see][for details]{2007A&A...463..957K} and hence not recognized as a separate component in this work.

To predict the internal detector background in a given energy band, we apply a simple approach of measuring detector count rate when \intl\ performs observations at high Galactic latitudes, where contribution from the GXB is negligible.

We divided the sky into two areas: \texttt{GAL} ($|l|<90^{\circ}$, $|b|<30^{\circ}$), which contains GXB, and \texttt{BKG} ($|l|\ge90^{\circ}$, $|b|\ge30^{\circ}$), used to trace the detector background. \texttt{GAL} and \texttt{BKG} sky regions contain 70226 and 61214 observations, with a total exposure of 112 and 114 Ms, respectively. 

The background model implements a simple approach based on tracking detector count rate during each \intl\ orbit (${\sim}3$ days). We assume that in-orbit variation of the background can be approximated by a linear law as a function of orbital phase $\Phi_{\rm orb}$ within the range 0.2$-$0.8. Depending on the number of available \texttt{BKG} {\scw}s ($N_{\rm BKG}$) within each orbit, we consider the following three options:
\begin{itemize}
    \item $N_{\rm BKG}$ $\geq$ 10 and the phase difference between the first and last \texttt{BKG} {\scw} is greater than 0.4: the background model is based on the linear regression $\alpha\times\Phi_{\rm orb}$+$\beta$, which also provides the uncertainty of the model.
    \item $3 < N_{\rm BKG} < 10$: the background model is a constant value, estimated as the mean detector count rate over the corresponding \texttt{BKG} {\scw}s. The error is calculated from a standard deviation. 
    \item $N_{\rm BKG} < 3$: The background model is calculated as a linear interpolation between two closest orbits with a background model calibrated by one of the two above cases. The uncertainty is also estimated by interpolation. Obviously, this option is triggered when no BKG {\scw}s are available during a given orbit. 
\end{itemize}

Some \intl\ orbits qualifying for the first of the above variants demonstrate rapid evolution of the detector count rate, usually due to the observation of a very bright X-ray source in outburst. Based on the distribution of the slope $\alpha$, we additionally filtered out 673 orbits by the condition $-2\times10^{-3}<\alpha<10^{-3}$ cts~s$^{-1}$~pix$^{-1}$, resulting in 68406 and 56996 of \texttt{GAL} and \texttt{BKG} {\scw}s with total exposures of 109 and 107 Ms, respectively.

Since the background model is based on high latitude observations (\texttt{BKG} region), its smooth work depends on how \texttt{BKG} {\scw}s are located among \texttt{GAL} ones. Ideally, \texttt{BKG} {\scw}s should be uniformly distributed among \texttt{GAL} data. For instance, the multi-year program of Galactic latitude scans (PI: Sunyaev) designed for GRXE study, combines Galactic and high-latitude observations during the same orbits in order to trace background variations. Unfortunately for this study, typical \intl\ observational programs are mainly concentrated within the Galactic plane without high-latitude background measurements. In the paradigm of the current background model, a number of \texttt{GAL} {\scw}s were made without representative nearby detector background, so that the GXB measurement is done with significant over- or under-estimation of the detector background. This usually manifests itself as a step-like behaviour in the residuals per orbit time interval. This effect increases the systematic noise and must be taken into account. 

To this end, we performed additional filtering of the data in the following way. We constructed a light curve of the residuals (observed minus predicted count rate) with a time bin of one orbit. Then, we accumulated the distribution of the residuals. The distribution is characterized by a narrow peak and wide wings. To suppress the wings, we applied a 3$\sigma$ iterative clipping algorithm. We performed this procedure over all data set (\texttt{GAL}+\texttt{BKG}) in the last energy interval of spectral binning (168$-$185~keV). At these energies the effective area of the \isg\ detector is small, and the detector count rate is expected to be dominated by the internal detector background. As a result, we excluded 231 orbits from the analysis. The final data set contains 58474 and 55619 of \texttt{GAL} and \texttt{BKG} {\scw}s with exposure of 94 and 105 Ms, respectively.

\begin{figure}
	\centerline{\includegraphics[width=\columnwidth]{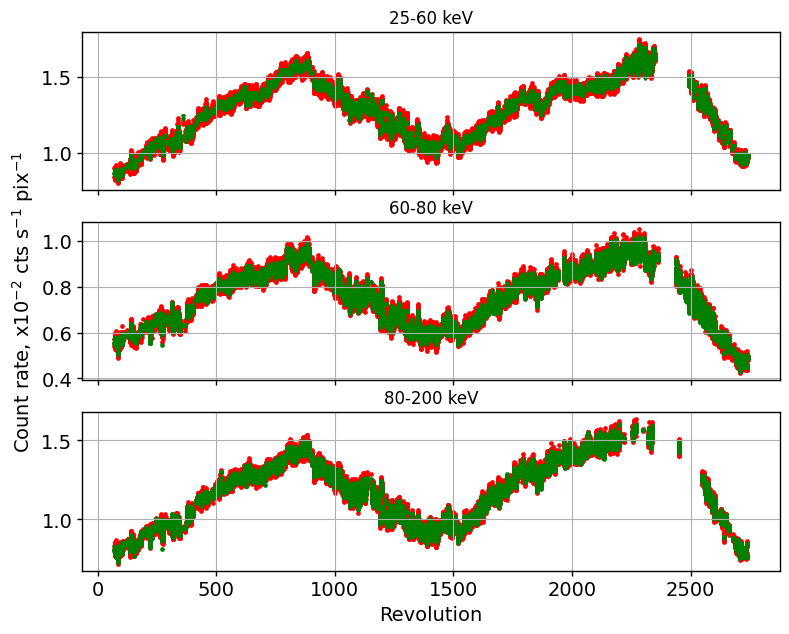}}
    \caption{\ib/\isg\ detector count rate from the \texttt{BKG} region (red points) and the count rate predicted by the background model (green points).}
    \label{fig:bkg:lc}
\end{figure}

\begin{figure}
	\includegraphics[width=\columnwidth]{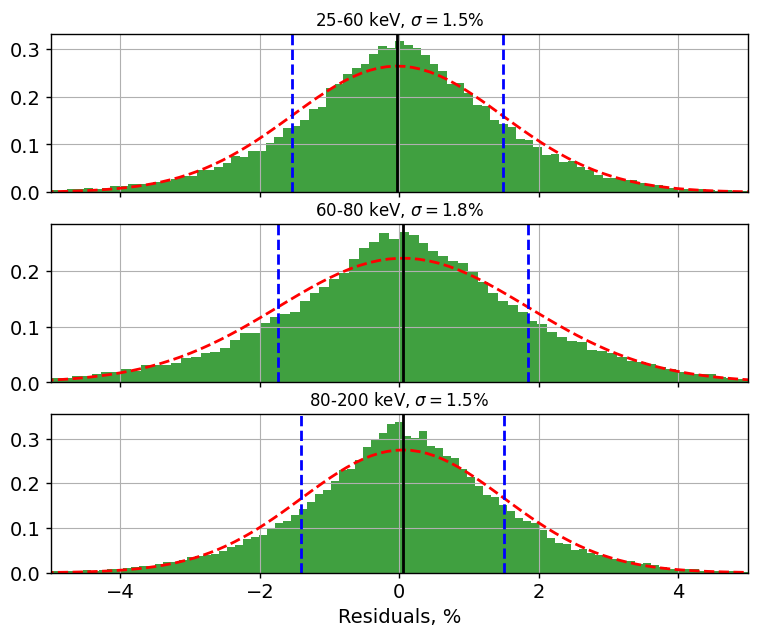}
    \caption{Normalized distribution of the relative residuals of the background model obtained in three energy bands (\texttt{BKG} region, the light curve is shown in Fig.~\ref{fig:bkg:lc}). The red dashed line shows the best-fitting Gaussian function. The vertical solid line and dashed lines represent the best-fitted mean $\mu$ and standard deviation $\sigma$ of the Gaussian function, respectively.}
    \label{fig:bkg:sys}
\end{figure}

\begin{figure}
	\includegraphics[width=\columnwidth]{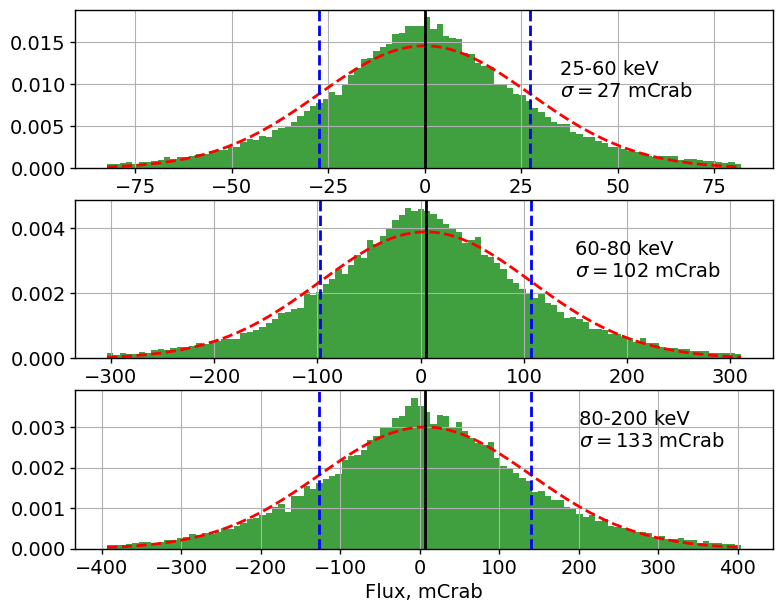}
    \caption{Similar to Fig.~\ref{fig:bkg:sys}, normalized distribution of the residuals expressed in absolute flux units (mCrab).}
    \label{fig:bkg:hist}
\end{figure}

In Fig.~\ref{fig:bkg:lc} we show \isg\ detector count rate in three energy bands extracted from the \texttt{BKG} sky region as a function of \intl\ orbit, along with the prediction of the background model. The noticeable two-humped variation is determined by the solar cycle, as also traced by the \intl\ $\gamma$-ray spectrometer SPI, by its anti-coincidence shield system \citep{2018A&A...611A..12D} and the count rate of the instrumental lines \citep{2022A&A...660A.130S}.

The distribution of residuals between the detector count rate and predicted background is shown in relative residuals (Fig.~\ref{fig:bkg:sys}) and in absolute flux (Fig.~\ref{fig:bkg:hist}) by using Crab calibration (Appendix~\ref{sec:calibr}). This distribution is characterized by a narrow peak and wide negative and positive wings, not described by a Gaussian function. Nevertheless, the residuals scatter around zero with a standard deviation of 27, 102, and 133 mCrab in the corresponding energy bands of 25$-$60, 60$-$80 and 80$-$200~keV. The relative accuracy ${<}2\%$ indicates a sufficiently well-described background model.

We measure the X-ray surface brightness in a given sky region by fitting the distribution of residuals described above with a Gaussian function. The position $\mu$ of the Gaussian provides the flux per {\ib} FOV. The uncertainty of the flux is calculated as $\sigma\times N_{\rm scw}^{-1/2}$, where $\sigma$ is the best-fitted width of the Gaussian, and $N_{\rm scw}$ is the number of {\intl} {\scw}s used. To take the   negative and positive wings into account separately, we added two constant functions defined below and above the centroid of the Gaussian ($\mu$). This is done to describe asymmetric wings that appear in some GXB measurements.

\section{\ib/\isg\ Crab calibration}
\label{sec:calibr}

\cite{2012A&A...537A..92K} estimated the effective solid angle of the IBIS telescope's FOV  $\Omega\approx286$~deg$^{2}$ by integrating the \ib/\isg\ collimator response function. The latter can be constructed by accumulating the shadowgram cast by Crab on the \isg\ detector as a function of angular offset. The \ib/\isg\ collimator response function is usually represented by a piece-wise linear function. The break position at ${\sim}4.5$~deg corresponds to the division between $9^{\circ}\times9^{\circ}$ fully coded and $29^{\circ}\times29^{\circ}$ partially coded FOV of the \ib\ telescope. 

In this work, we measure the intensity of the GXB as an excess above the detector internal background within the effective solid angle $\Omega$. The excess, expressed in terms of count rate (\cnts) in a given energy band, is converted to physical units (\flux) using the observed count rate of the Crab Nebula, a bright and stable X-ray source widely used in X-ray astronomy as a standard candle. The Crab spectrum can be approximated by a power law, $dN/dE = N\,E^{-\Gamma}$ photon cm$^{-2}$\,s$^{-1}$\,keV$^{-1}$. We adopt $N=10$ and $\Gamma=2.1$ to be compatible with our previous works. 

%however recent Crab measurements are available \citep[e.g.,][]{2017ApJ...841...56M}. \textcolor{red}{What is this last phrase for? If we do not use these recent measurements, then let's remove the second half of the sentence.}

\begin{figure}
	\includegraphics[width=\columnwidth]{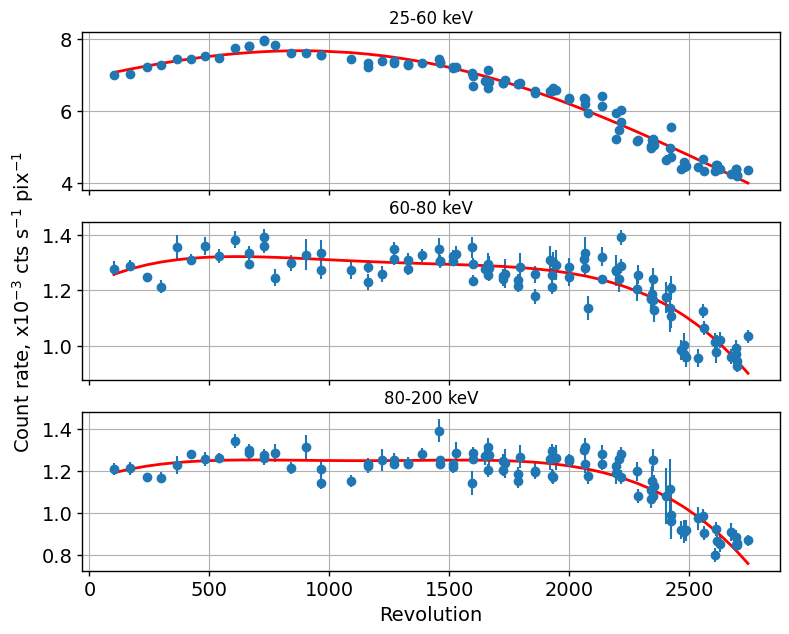}
    \caption{Observed \ib/\isg\ detector count rate of the Crab Nebula (points). The red line represents a cubic polynomial approximation.}
    \label{fig:crabmodel:poly}
\end{figure}

Since we analyze multi-year {\intl} data, the observed count rate of the Crab in a given energy band is subject to change due the long-term evolution  of the detector efficiency. To trace this evolution, we selected {\intl} orbits for which at least ten Crab {\scw}s were carried out at angular distance less than 4.5 degrees. Fig.~\ref{fig:crabmodel:poly} shows the observed Crab count rate as a function of {\intl} revolution for the 25$-$60, 60$-$80 and 80$-$200~keV bands. Note that the current {\isg} calibration is available until the year 2020 (${\sim}2250$ {\intl} revolution) and the strong decline in the Crab count rate after this epoch could be due to aging of the instrument and extrapolation of the calibration. The Crab count rate also shows some high variations and spikes, which we attribute to systematic noise.

Assuming that the detector efficiency varies smoothly, we approximated the observed count rate with a cubic polynomial fit. We have verified that changing the fitting parameters does not significantly affect the results. The resulting function is used to convert the observed detector count rate in a given energy band for any {\intl} {\scw} into Crab units, and then, to physical flux, using close in time Crab observations. This continuous re-calibration  allows us to mitigate the aging of the instrument and inaccuracies in the {\isg} calibration. This method also provides a smooth recalibration of the {\ib}/{\isg} effective area (or Ancillary Response File, ARF) over the time span of the mission.

\begin{figure}
	\includegraphics[width=\columnwidth]{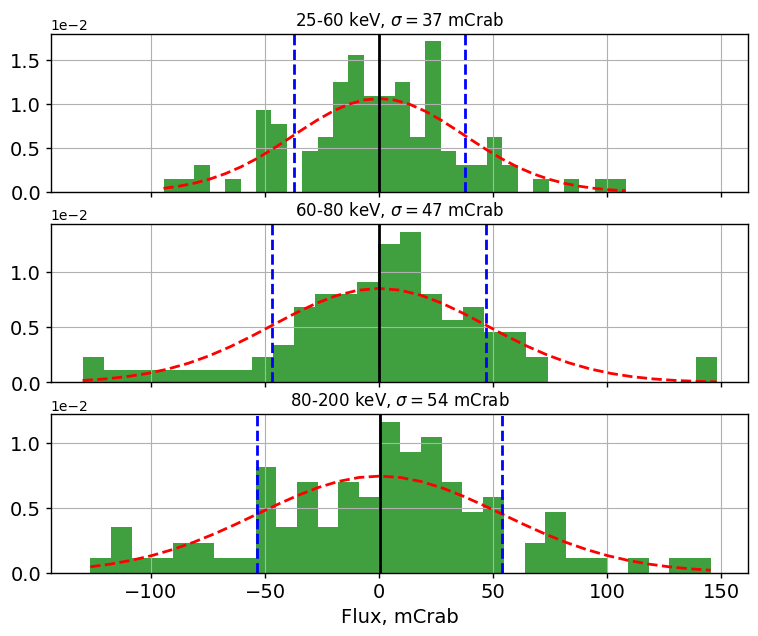}
    \caption{Normalized distribution of the residuals between the \ib/\isg\ Crab count rate (Fig.~\ref{fig:crabmodel:poly}) and the corresponding polynomial fit. The distribution is approximated with a Gaussian function. The vertical black solid and green dashed lines represent the centroid of the distribution (consistent with zero) and standard deviation, respectively.}
    \label{fig:crabmodel:sys}
\end{figure}

To estimate the systematic uncertainty of this procedure, we accumulated the distribution of the residuals between the observed Crab count rate and polynomial fit for a given orbit in mCrab units, and approximated it with a Gaussian (see Fig.~\ref{fig:crabmodel:sys}). The standard deviation ${\sigma}=37$, 47 and 44 mCrab, respectively, for the 25$-$60, 60$-$80 and 80$-$200~keV bands is adopted as a systematic uncertainty of the GRXE flux measurement. Note that it exceeds by at least an order of magnitude the statistical uncertainty estimated from the photon statistics. For this reason we decided to ignore the statistical errors for GRXE measurements. 

To summarize, the ratio between the \isg\ detector count rate at every every moment of time in a given energy band and the corresponding polynomial value provides the flux in Crab units measured within the solid angle subtended by the \ib\ FOV.

\section{A test of the method using Sco~X-1 data}
\label{sec:scox1}

\begin{figure}
	\includegraphics[width=\columnwidth]{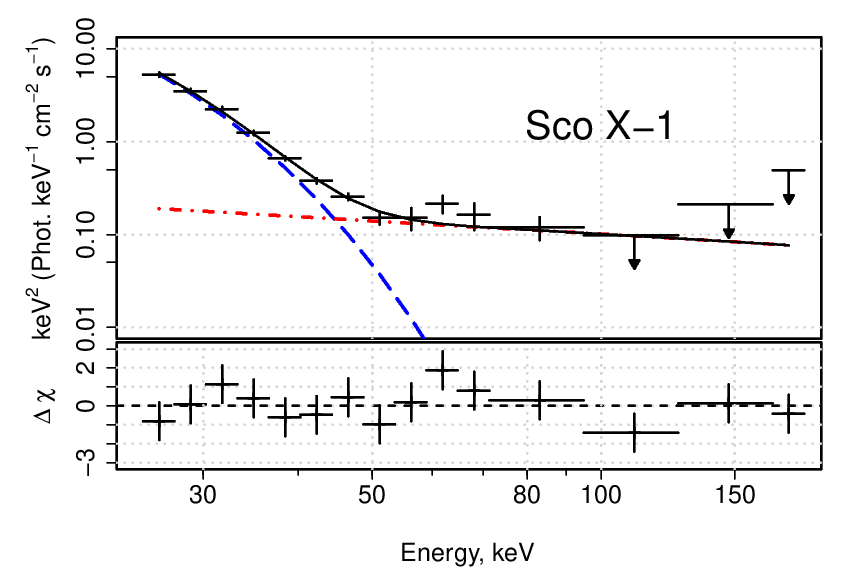}
    \caption{Time-averaged spectrum of Sco~X-1 obtained with {\ib}/{\isg} in collimator mode (black data points). The black line shows the best-fitting model with two components: simple analytic model $dN/dE\propto \exp(-E/3.9\textrm{~keV})$ and a power law with $\Gamma=2.5$, represented by the blue dashed and red dash-dotted lines, respectively.}
    \label{fig:scox1}
\end{figure}

To test our method, we extracted the spectrum of the bright point X-ray source Sco~X-1 using the same collimator mode as we use for the GXB in this study. To this end, we first selected 838 {\scw}s with angular separation of less than $4.5$~deg from the position of Sco~X-1, which adds up to 1.7~Ms of dead-time corrected exposure. We then normalized the {\isg} detector count rate of Sco~X-1 by the Crab count rate model (\ref{sec:calibr}), calculating the calibrated flux in mCrab units. Finally, using all available {\scw}s, we constructed the spectrum of Sco~X-1 shown in Fig.~\ref{fig:scox1}. 

Similar to \cite{2014MNRAS.445.1205R}, who analyzed the {\ib}/{\isg} spectrum of Sco~X-1 in the standard point-like mode, we approximated the obtained spectrum with a combination of (i) a power law with a fixed slope $\Gamma=0$ and a free high-energy cutoff energy $E_{\rm cut}$, representing the emission of the boundary layer on a neutron star surface, and (ii) a power law tail with a free slope $\Gamma$. The normalizations of both components were also free parameters. The original spectrum was fitted with a statistical quality of $\chi^{2}_{\rm r}$/dof = $1.62/17$. Similar to \cite{2014MNRAS.445.1205R}, we added a systematic uncertainty of 5\%, which improved the fit quality to $\chi^{2}_{\rm r}$/dof = $0.98/17$. The resulting model parameters: $E_{\rm cut}=3.93\pm0.22$~keV and $\Gamma=2.48_{-0.65}^{+0.79}$. \cite{2014MNRAS.445.1205R} found the spectrum of Sco~X-1 in different patterns of flux variations to be consistent with $E_{\rm cut}=3.70\pm0.03$~keV and $\Gamma=2.6\pm0.3$, in agreement with our findings within the uncertainties. 

%Note that the Sco~X-1 spectrum in collimator mode shows a roll-off at ${\sim}50$~keV, whereas the GXB spectrum reveals a minimum at ${\sim}80$~keV.} \textcolor{red}{What is this sentence for? Let's remove it}

We thus conclude that our method of using {\ib}/{\isg} as a collimator instrument provides reliable results, consistent with standard {\ib} sky deconvolution.

%% If you have bibdatabase file and want bibtex to generate the
%% bibitems, please use
%%
\bibliographystyle{elsarticle-harv} 
\bibliography{biblio}

%% else use the following coding to input the bibitems directly in the
%% TeX file.

%%\begin{thebibliography}{00}

%% \bibitem[Author(year)]{label}
%% For example:

%% \bibitem[Aladro et al.(2015)]{Aladro15} Aladro, R., Martín, S., Riquelme, D., et al. 2015, \aas, 579, A101

%%\end{thebibliography}

\end{document}